\documentclass{aa}  

\usepackage{graphicx}
\usepackage{txfonts}
\usepackage{hyperref}
\newcommand{\dg}{^{\circ}}

\begin{document}

   \title{Global alignments of parsec-scale AGN radio jets and their polarization planes}
   
   \titlerunning{Global alignments of parsec-scale AGN radio jets}

   \subtitle{}

   \author{D. Blinov
          \inst{1,2,3}
          \and
          C. Casadio\inst{4}
          \and
          N. Mandarakas\inst{1,2}
          \and
          E. Angelakis\inst{5,4}
          }

   \institute{Foundation for Research and Technology - Hellas, IESL \& Institute of Astrophysics, Voutes, 7110 Heraklion, Greece\\
              \email{blinov@ia.forth.gr}
         \and
             Department of Physics, University of Crete, 71003, Heraklion, Greece
        \and
             Astronomical Institute, St. Petersburg State University, Universitetsky pr. 28, Petrodvoretz, 198504 St. Petersburg, Russia
        \and
             Max-Planck-Institut f\"{u}r Radioastronomie, Auf dem H\"{u}gel 69, 53121 Bonn, Germany
        \and
             Section of Astrophysics, Astronomy \& Mechanics, Department of Physics, National and Kapodistrian University of Athens,\\ Panepistimiopolis Zografos 15784, Greece
             }

   \date{Received January 4, 2020; accepted }

 
  \abstract
   {A number of works have reported that the polarization plane of extragalactic sources as well as the structural axes of radio sources are aligned on the large scale. However, both the claims and their interpretation remain controversial.}
   {For the first time, we explore the alignment of parsec-scale jets. 
   Additionally, we use archival polarimetric data at different wavelengths in order to compare the relative orientations of the jets and the polarization planes of their emission.}
   {Using the flux density distribution in very long baseline interferometry (VLBI) radio maps from the Astrogeo database, we determine the parsec-scale jet orientation for the largest sample of active galactic nuclei (AGN) to date. Employing the method of parallel transport and a sample statistics characterizing the jet orientation dispersion among neighbors, we test whether the identified jets are significantly aligned.}
   {We show that the parsec-scale jets in our sample do not demonstrate any significant global alignments. Moreover, the jet direction is found to be weakly correlated with the direction of the polarization plane at different frequencies.}
   {}


   \maketitle
%

\section{Introduction}

Finding the large-scale structures in the Universe is of great interest for cosmology. 
Characterizing of such structures allows us to place stringent constraints on cosmological models 
and their parameters \cite[e.g.,][]{Multamaki2004,Pavlidou2014}. The large-scale structures can be 
identified either via detection of objects clustering \citep{Gott2005,Clowes2013,Balazs2015} or 
using the coherence of their characteristics. For instance, there have been attempts to investigate 
possible global alignments of the optical and radio polarization planes of active galactic nuclei 
(AGN) as well as alignments of their morphological axes.

\cite{Hutsemekers1998} demonstrated that the optical polarization planes of a sample of 170 quasars 
are coherently oriented at very large spatial scales. This effect was later confirmed at higher 
significance levels with larger samples \citep{Hutsemekers2001,Hutsemekers2005} and independently 
using different methods \citep{Jain2004}.

A number of studies has later been conducted in order to search for similar alignments in much 
larger samples of AGN with measured radio polarization. \cite{Joshi2007} analyzed the polarization 
angles of 4290 sources from the 8.4 GHz Jodrell/VLA Astrometric Survey (JVAS) and Cosmic Lens 
All-Sky Survey (CLASS) data \citep{Jackson2007} and found no systematic alignments. They also 
determined jet position angles from VLBI observations of 1565 sources and tested this sample for 
global alignments. No significant signal was detected in this sample either. However, 
\cite{Tiwari2013} and \cite{Pelgrims2016} used largely the same JVAS and CLASS data and presented 
evidence for the global alignment of polarization vectors in their sample. Moreover, 
\cite{Pelgrims2016} demonstrated that the radio polarization position angles of the quasars are 
aligned with the axes of the large quasar groups they belong to. \cite{Tiwari2019} analyzed 
high-frequency 86 and 229 GHz polarization measurements of 211 AGN from \cite{Agudo2014}. They found 
that the polarization orientations are consistent with the assumption of isotropy at scales larger 
than or equal to $\sim$800 Mpc.

Several studies have reported that morphological structures of radio sources are aligned at 
cosmological scales. \cite{Jagannathan2014} used a survey of the ELAIS N1 field conducted at 615 MHz 
with the Giant Meter-wave Radio Telescope (GMRT) and identified a sample of 65 sources with extended 
radio jets. They found that the directions of these jets are correlated at angular separation scales 
up to 1.8$\dg$, which corresponds to 53 Mpc at a redshift of 1. \cite{Contigiani2017} analyzed a 
sample of position angles corresponding to preferential directions of 30059 radio sources. This 
sample was created within the Radio Galaxy Zoo project using images of the Faint Images of the Radio 
Sky at Twenty-centimeters survey \citep{Becker1995}. They found that the position angles on angular 
scales 1.5$\dg$ - 2$\dg$ are aligned at a 3.2$\sigma$ significance level. This angular scale roughly 
corresponds to a spatial scale of 19 - 39 Mpc for the mean redshift in their sample. At the same 
time, \cite{Contigiani2017} did not detect any similar alignment in a sample of 11674 extended 
sources from the TIFR GMRT Sky Survey \citep{Intema2017}. This was attributed to the sparsity of 
this sample.

Global alignments of optical polarization vectors at large angular scales can be expected either as
a result of propagation effects or by coherent orientation of extragalactic sources. For instance, 
the Galactic magnetic field produces a large-scale alignment of the dust particles in the 
interstellar medium, which preferentially attenuates a particular polarization component of the 
electromagnetic waves \cite[e.g.,][]{Heiles1996}. This means that if the intrinsic polarization of 
extragalactic sources is low, the interstellar polarization may give a rise to optical polarization 
that is aligned over large distances over the sky as determined by the size of the intervening 
interstellar dust clouds. However, current studies are typically aware of this effect, and so far, 
no evidence of a significant contribution to the polarization alignments has been confirmed 
\citep{Pelgrims2019}. Moreover, the Galactic extinction cannot explain the reported redshift 
dependence of the polarization alignments \citep{Hutsemekers2005}. Another effect that might modify 
the polarization of light along the line of sight is the mixing of photons with axion-like particles 
\citep{Das2005}. This effect is expected to produce high circular polarization comparable to the 
level of linear polarization \citep{Hutsemekers2011}. However, because high levels of circular 
polarization have never been observed in extragalactic sources 
\citep[e.g.,][]{Takalo1993,Hutsemekers2010}, this mechanism has essentially been ruled out. 
Supported by the reported alignments of the morphological structure of radio sources, the second 
scenario where the polarization alignment is caused by the coherent orientation of extragalactic 
sources therefore appears to be more favorable.

\cite{Hutsemekers2014} found that AGN polarization vectors are either parallel or perpendicular to  
the large-scale structures (large quasar groups) to which they belong, and thereby the polarization 
vectors of neighboring sources are correlated. Under the assumption that AGN polarization is either 
parallel or perpendicular to the accretion disk axis, they inferred that quasar spin axes are likely 
parallel to their host large-scale structures. This conclusion is in agreement with the 
observations showing alignments of galactic spin axes with the cosmic web 
\citep{Tempel2013,Jones2010}, and modeling results that showed the coevolution of galaxies with 
their host large-scale structures \citep{Catelan2001,Wang2018}.

In this paper we obtain and study the largest set of 6388 parsec-scale AGN radio jet position angles 
to date. The aim of the work is twofold: (1) we characterize the relation of these jet directions 
to the polarization of sources at different frequencies and examine whether the polarization can 
serve as a proxy of the jet orientation, and (2) we investigate whether the jet position angles are 
aligned for sources that are separated by large angular distances on the sky.

Throughout the paper we use the IAU convention for position angles (PA) and the electric vector  
position angles (EVPA), where they are calculated from north to east. The values of the cosmological 
parameters adopted throughout this work are $H_0 = 67.8$ km s$^{-1}$ Mpc$^{-1}$, $\Omega_m = 0.308$, 
and $\Omega_\Lambda = 1 - \Omega_m$ \citep{Planck2016}.

\section{Data sample and reduction} \label{sec:data}
\subsection{Jet directions} \label{subsec:jet_dir}
\begin{figure*}
   \centering
   \includegraphics[width=0.35\textwidth]{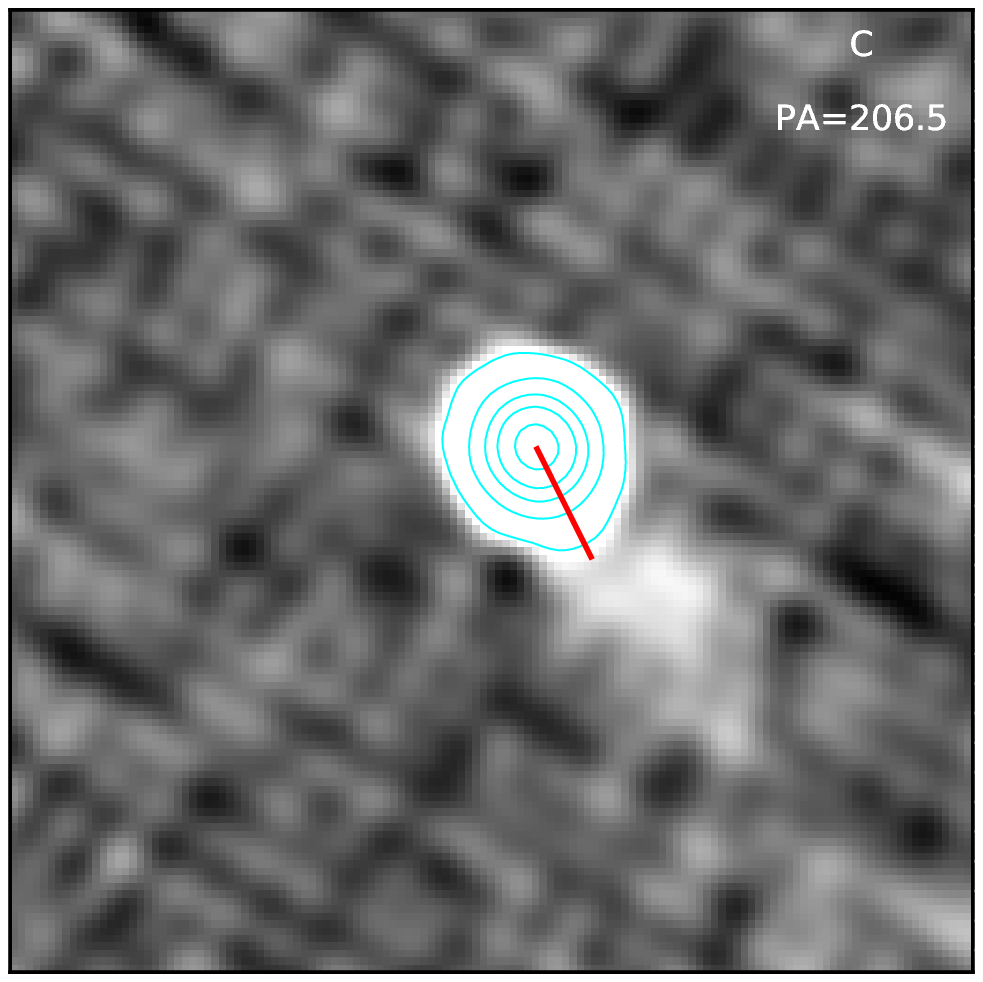}
   \includegraphics[width=0.35\textwidth]{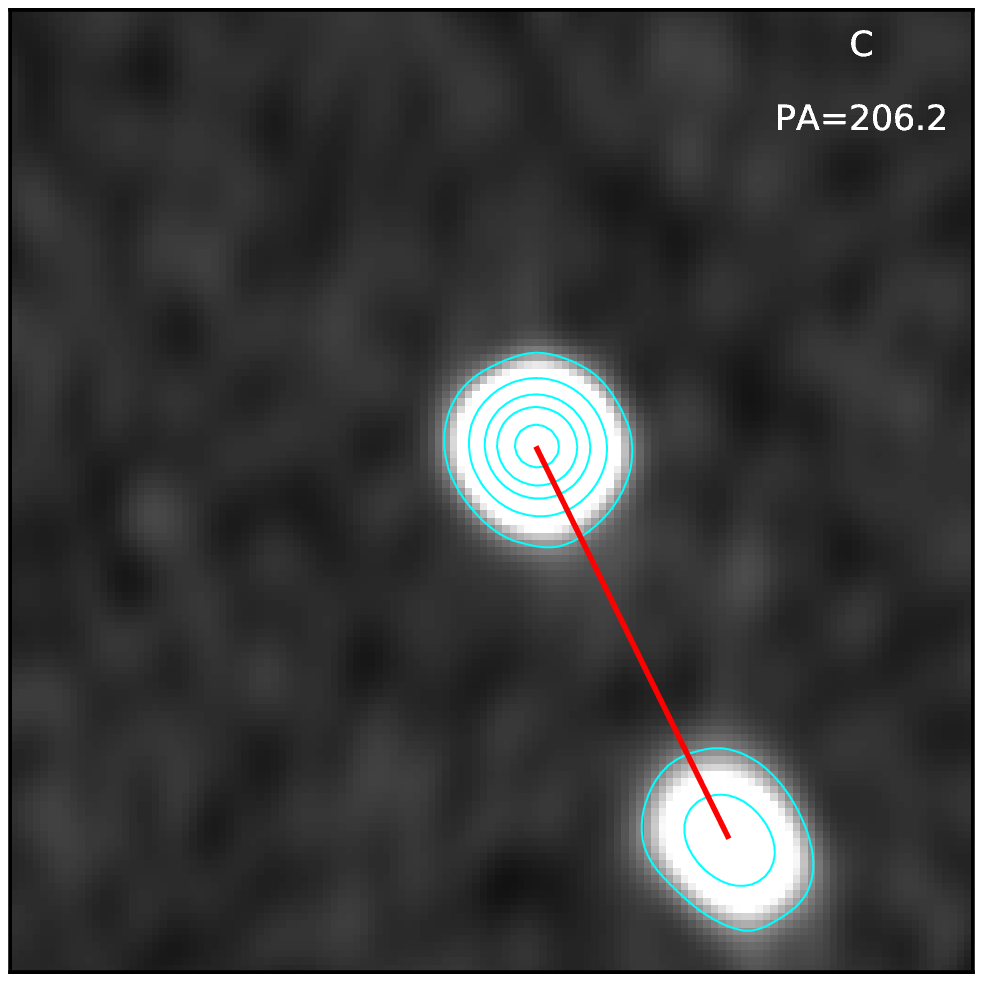}
   \includegraphics[width=0.35\textwidth]{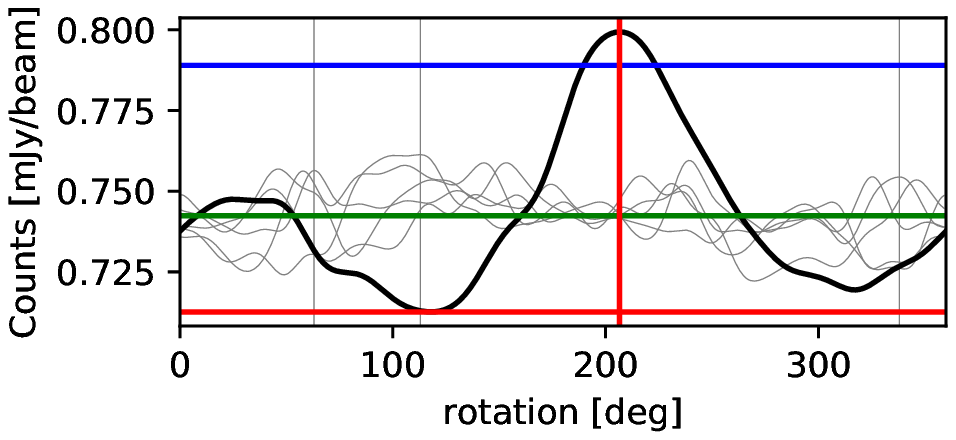}
   \includegraphics[width=0.35\textwidth]{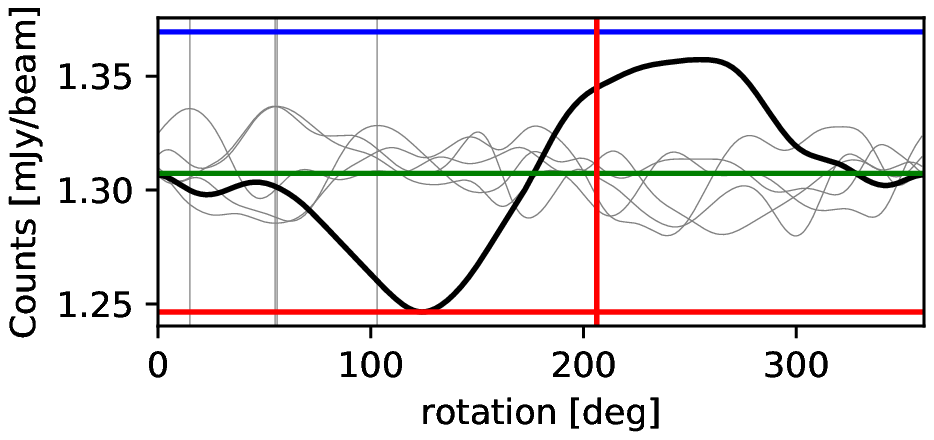}
    \caption{Examples of the jet detection. The left panel shows the C-band image of J0933+6106 
(RA=09h33m10.4s, DEC=+61d06m46s) (top) and its azimuthal flux density distribution within the 
$3.5\sigma$ circular area (bottom). The right panel presents similar plots for J1005+2403 
(RA=10h05m07.9s, DEC=+24d03m38s). The red lines indicate the determined PA$_{\rm jet}$. The solid 
black line shows the azimuthal brightness distribution for the source, and five gray lines show the 
same value calculated in random positions within 40 mas from the source. Both images are from the 
VLBA Imaging and Polarimetry Survey \citep[VIPS,][]{Helmboldt2007}.}
    \label{fig:alg}
\end{figure*}

We used data from the Astrogeo VLBI FITS image database\footnote{\url{http://astrogeo.org/vlbi_images/}}, which contains images of compact radio sources, mainly AGN, that have been obtained by various VLBI experiments. We obtained the Astrogeo data set version on 30 March 2019, when it contained 92220 radio maps of 14078 compact radio sources. Below we describe the procedure that was followed to  determine the jet direction, PA$_{\rm jet}$, for these sources.

The original images in the Astrogeo database are convolved with elliptical beams. Because our jet determination procedure operates in the image plane, the beam elongation could bias the results \citep{Pushkarev2017}. Therefore we reimaged every source by convolving it with a median (if multiple epochs were present) circular beam calculated as $r = \sqrt{a b}$, where $a$ and $b$ are the dimensions of the elliptical beam. Subsequently, images of different epochs were aligned such that the maximum pixel was at the image central pixel, and then they were median stacked.

In the stacked images we fit the main (central) source component with a circular Gaussian. The standard deviation of this Gaussian, $\sigma$, defined a stripe of $3.5\, \sigma$ pixels in length and 2 pixels in width starting at the central pixel and pointing towards the north pole of the celestial frame. Rotating the stacked image with steps of $0.5\dg$ with interpolation 720 times and summing the counts within the stripe, we obtained the azimuthal distribution of the surface brightness for each source and frequency. Two examples of such distributions are shown in Fig.~\ref{fig:alg}. When the maximum of the flux density profile distribution was 2.4 times its standard deviation, the jet was considered detected. Its position angle was then defined by the image rotation angle corresponding to the maximum (left column of Fig.~\ref{fig:alg}). When the 2.4 threshold was not exceeded by the maximum of the flux density distribution (right column of Fig.~\ref{fig:alg}), a different procedure of the jet direction determination was called. First, the circular area within $3.5\sigma$ around the center was masked. Then the flux density maximum of the remaining image was determined. When this peak was $3 \sigma$ above the surrounding background, another Gaussian centered at this secondary peak was fit to the image. If the latter fitting procedure converged and the Gaussian width was in the range $0.6$ mas $ < \sigma < 5$ mas, then the jet position angle was considered to be equal to the Gaussian peak position angle. This approach was adopted in order to handle cases with a symmetric core and a bright isolated knot in the jet, similar to the case of J1005+2403 shown in the right panel of Fig.~\ref{fig:alg}. All the parameters used in both procedures were determined empirically with the requirement to maximize the jet detection efficiency.

We found that the algorithm described above is faster and gives a more reliable determination of the internal jet position angle than the ridge-line determination and smoothing using slices within concentric annuli. Caveats related to the algorithm, jet bending, and frequency dependence of position angles are discussed in Sec.~\ref{sec:res}.

Results of the automatic jet position angle detection procedures were verified visually by the authors. To this end, we created a web service  that is conceptually similar to interfaces of citizen-science projects  \citep[e.g.,][]{Banfield2015}. At every web-page reload a user could see the results of the jet detection algorithm for a single random source similar to those shown in Fig.~\ref{fig:alg}. The randomization was introduced to avoid possible biases. Plots for all available bands were shown together with corresponding stacked fits images. The latter were displayed in
JS9\footnote{\url{https://js9.si.edu/}} windows, which allows changing the dynamic range levels, the zoom, the scale, etc. Users judged the resulting jet position angles in plots with a click at one of four buttons "good", "bad", "no jet" and "unclear" after inspecting the images corresponding to every band present for the source. Each source in the database was inspected by at least one of the authors, and the corresponding choice was recorded. The database with results contains 27629 entries in total. It contains 9822 images in which the jet direction is identified by one of the algorithms that is conformed with the visually identified jet (labeled "good"); 2478 where the calculated jet direction does not agree with the visual jet ("bad"); 13520 images where the jet is not seen ("no jet") and 1809 images where it is unclear whether the jet is real or is an artifact of data reduction ("unclear"). For the further analysis we conservatively use only the sources or bands for which both machine and human identified jets directions agree, which is those labeled "good".

\subsection{Radio and optical polarization} \label{subsec:rad_pol}

We collected archival data of optical and radio polarization for large samples of AGN.
The polarization plane orientation for 12746 sources at 8.4 GHz was taken from \cite{Jackson2007}, 
where polarimetric data of JVAS \citep{Patnaik1992} and CLASS \citep{Myers2003} surveys were 
uniformly combined. In our analysis we used only the 7095 measurements for which the uncertainty of 
EVPA was lower than $15\dg$.

The polarization plane directions for 183 sources at 86 GHz was taken from \cite{Agudo2014}. The 
average EVPA uncertainty in this sample was $3.5\dg$ and does not exceed $11\dg$. Therefore we did 
not use any additional significance cuts.

Optical polarization measurements were collected from 
\citet{Hutsemekers2005,Hutsemekers2018,Itoh2016}  and \citet{Angelakis2016}, where the measurements 
were performed in V and R bands. The first two works contain single-epoch measurements. Multiple 
measurements of a source from the later two papers were averaged by finding centroids of the 
distributions on the Q - U Stokes parameter plane. In the analysis, we used only the 406 
measurements from the collected sample where the uncertainty of the EVPA was lower than $15\dg$.

\section{Results} \label{sec:res}

\begin{figure}
   \centering
   \includegraphics[width=0.45\textwidth]{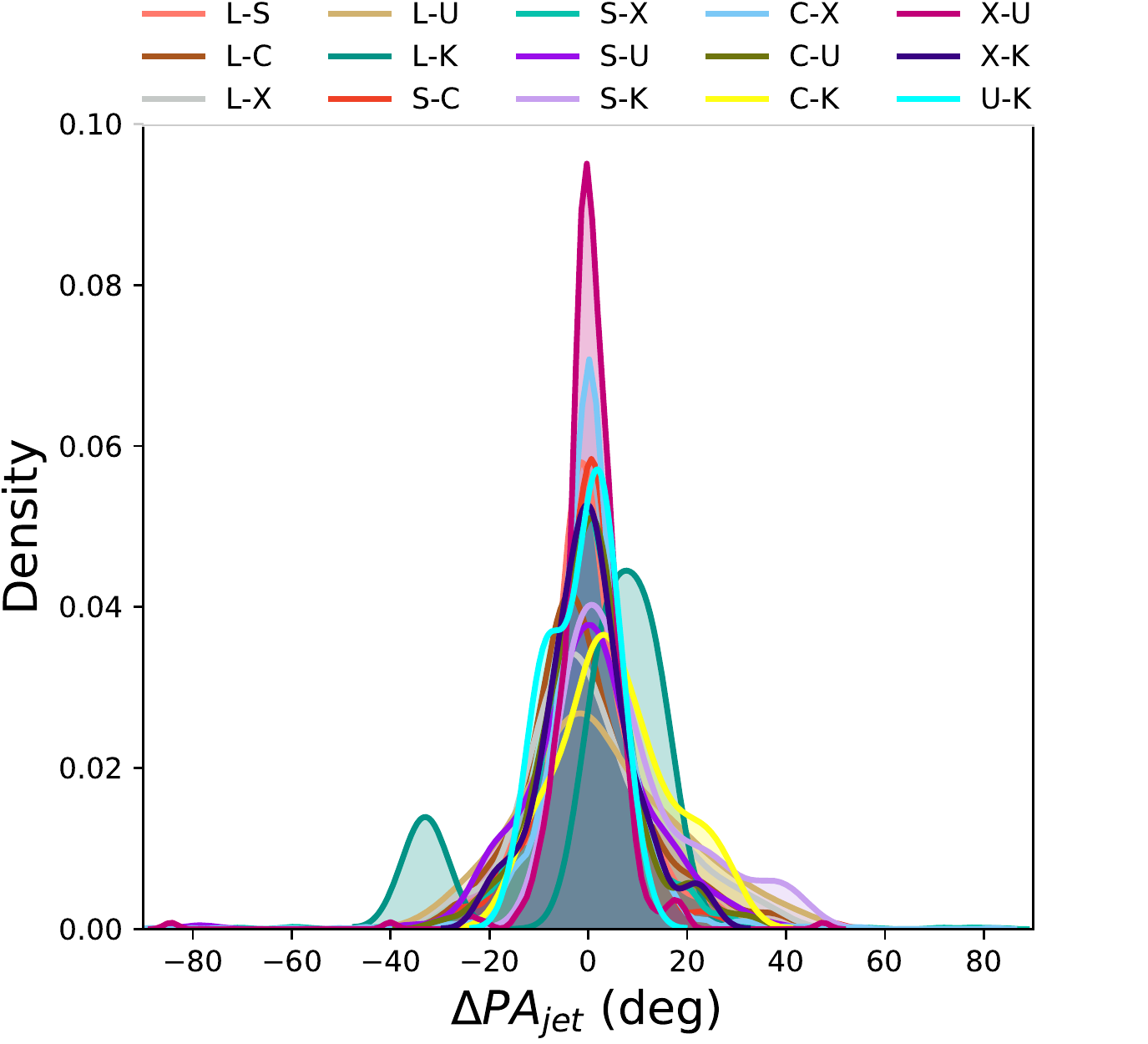}
    \caption{Distribution of pairwise differences of detected jet directions at different frequencies.}
    \label{fig:freq_comp}
\end{figure}

After the visual verification, we obtained confident jet direction estimates for a total of 6388 sources. For 2723 sources the jet position angle was available at more than one frequency. The jet position angle is known to be variable in time \citep{Lister2013}. Because we stacked data for different epochs when available, this variability was partially averaged. Furthermore, the jet at parsec  and kiloparsec scales is not always parallel \citep{Kharb2010}, and moreover, even at the parsec (milliarcsecond) scale,  significant jet bending is not uncommon \citep[e.g.,][]{Rastorgueva2009}. In order to estimate uncertainties in the derived inner PA$_{\rm jet}$ , we calculated pairwise differences of their values at separate frequencies and constructed distributions of these differences. These distributions are shown in Fig.~\ref{fig:freq_comp}, and their standard deviations are listed in Table~\ref{tab:pa_diff}. The two highest frequency bands Q (43 GHz) and W (86 GHz) are not presented there because there is only one source with PA$_{\rm jet}$ determined in each of them. The standard deviation of PA$_{\rm jet}$ differences between bands does not exceed $16\dg$. Moreover, for neighboring (in frequency) bands, it is $<10\dg$ in all cases. Therefore, this value can be considered as an estimate of the uncertainty of the PA$_{\rm jet}$ determination in our data set.

\begin{table}
\caption{Standard deviation of the distributions of pairwise differences in derived jet position angles measured in degrees.}
\label{tab:pa_diff}
\begin{tabular}{lcccccc}
\hline
Band        & L    & S    & C    &  X   & U    & K \\
Freq. (GHz) & 1.4  & 2.3  & 4.3  & 8.4  & 15.4 & 24.4 \\
\hline
L           &   -  & 9.2  & 12.1 & 13.6 & 15.2 & 15.9 \\
S           & 9.2  &   -  & 9.7  & 12.0 & 13.1 & 12.4 \\
C           & 12.1 & 9.7  &   -  & 9.7  & 10.0 & 10.8 \\
X           & 13.6 & 12.0 & 9.7  &   -  & 7.9  & 8.8  \\
U           & 15.2 & 13.1 & 10.0 & 7.9  &   -  & 5.9  \\
K           & 15.9 & 12.4 & 10.8 & 8.8  & 5.9  &  -   \\
\hline
\end{tabular}
\end{table}

As a result of jet bending and opacity effects \citep{Kovalev2008} , observations at higher frequencies give a better approximation of the innermost jet direction. For each source with PA$_{\rm jet}$ determined in more than one band, we therefore selected the value corresponding to the highest available frequency. The number of sources with PA$_{\rm jet}$ determined at each particular band is listed in Table~\ref{tab:n_freq}.

\begin{table}
\centering
\caption{Distribution of 6388 sources in the sample over frequencies where the PA$_{\rm jet}$ was determined.}
\label{tab:n_freq}
\begin{tabular}{lcc}
\hline
Band        & Freq. (GHz) & N \\
\hline
L           & 1.4         & 3    \\
S           & 2.3         & 567  \\
C           & 4.3         & 1591 \\
X           & 8.4         & 3721 \\
U           & 15.4        & 481  \\
K           & 24.4        & 23   \\
Q           & 43.3        & 1    \\
W           & 86.2        & 1    \\
\hline
\end{tabular}
\end{table}

\begin{figure}
   \centering
   \includegraphics[width=0.42\textwidth]{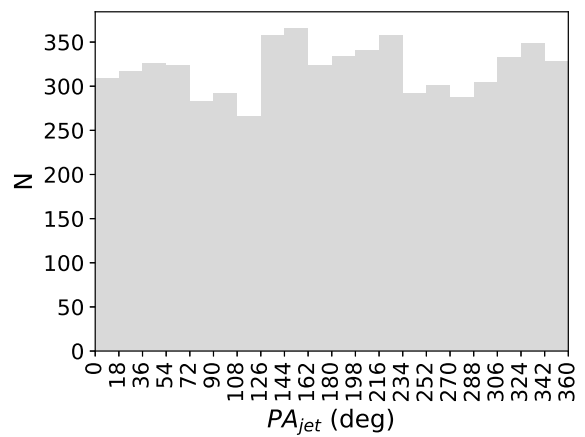}
    \caption{Distribution of the jets position angles PA$_{\rm jet}$.}
    \label{fig:pa_hist}
\end{figure}

Figure~\ref{fig:pa_hist} shows the distribution of PA$_{\rm jet}$. There are two visible minima in 
the histogram around $90\dg$ and $270\dg$. However, according to the Kolmogorov-Smirnov test the 
distribution of PA$_{\rm jet}$ is consistent with the uniform distribution ($p$-value=0.018). This 
pattern is possibly caused by the geometric configuration of the array, mainly Very Long Baseline 
Array (VLBA) and European antennas, involved in the observations presented in the Astrogeo database. 
Similar nonuniformity was found by \cite{Contigiani2017}, who obtained a triple-peaked pattern 
associated with the shape of the Very Large Array (VLA) and possibly to the side-lobe structure 
created by it \cite{Helfand2015}. Because the nonuniformity is marginal and is not a local effect, 
this does not affect our analysis \citep{Contigiani2017}.

We are interested in the jet axis orientation rather than the detected jet vector direction, therefore we reduced the PA$_{\rm jet}$ value range in the following analysis from $[0\dg,360\dg)$ to $[0\dg,180\dg)$.

\section{Jet versus polarization} \label{sec:pa_vs_evpa}

Most of the previous studies dealt with global alignments of the EVPA in optical or radio emission of sources. As we discussed in the introduction, the interpretation of the results is not straightforward in these cases because the polarization plane may rotate as the electromagnetic wave propagates. The observation of polarization alignment can therefore imply either the intrinsic alignment of polarization and morphological axes of sources in the absence of propagation-induced rotations, or propagation effects that force the alignment of otherwise intrinsically uncorrelated polarizations such as the absorption-induced polarization. In the first case the orientation of the polarization plane is considered to be a proxy of the orientation of the structural axes (PA$_{\rm jet}$ in our case) \citep[e.g.,][]{Rusk1985,Lister2000}. Having in hand the largest data set of jet position angles to date, we test whether they are indeed correlated with the radio and optical EVPAs.

\subsection{Jet direction versus radio EVPA at 8 GHz} \label{subsec:comp_8ghz}

\begin{figure}
   \centering
   \includegraphics[width=0.35\textwidth]{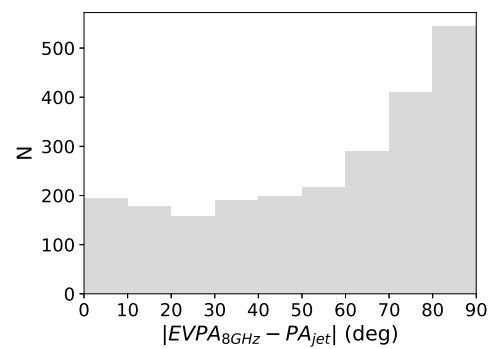}
      \caption{Distribution of the difference between the jet direction PA$_{\rm jet}$ and polarization position angle from the JVAS/CLASS 8.4GHz catalog.}
         \label{fig:jd_vs_8ghz}
\end{figure}
We cross-matched sources in our sample with the subsample of the JVAS/CLASS catalog of radio polarization at 8.4 GHz (see Sec.~\ref{subsec:rad_pol}). We found 2378 common sources separated by $< 1$ arcsec. The distribution of the difference between the polarization position angle and PA$_{\rm jet}$ of these sources is shown in Fig.~\ref{fig:jd_vs_8ghz}. The distribution shows a prominent peak at $90\dg$. However, the distribution has a wide spread, and only 40\% of the sources are located within $20\dg$ from the peak. In addition to the possibility that intrinsically emitted EVPA may not be ideally aligned with the jet orientation \citep{Peirson2019}, two major effects probably affect the distribution in Fig.~\ref{fig:jd_vs_8ghz}.

Partially, the spread of $|EVPA_{\rm 8\,GHz}-PA_{\rm jet}|$ may be caused by the Faraday rotation effect, which alters the emitted EVPA. Rotation measure (RM) values between 8 and 15 GHz range from a few 10$^{2}$ rad $\rm m^{-2}$ to roughly 10$^{4}$ rad $\rm m^{-2}$ in AGN jets \citep{Zavala2004, Kravchenko2017}. When we consider a median RM value between 8 and 15 GHz of 400 rad $\rm m^{-2}$, this would rotate the 8 GHz polarization planes by $\sim30\dg$ according to results in \cite{Hovatta2012}. It is therefore expected that the information on the emitted EVPA is completely lost for a significant fraction of sources at 8 GHz.

Another reason that likely contributes to the spread in Fig.~\ref{fig:jd_vs_8ghz} is that a significant fraction of PA$_{\rm jet}$ were obtained from images at different frequencies (see Table~\ref{tab:n_freq}), while a portion of the jet that contributes to the total flux density and integrated polarization decreases with frequency. At 8 GHz we are more sensitive to the extended emission than at higher frequencies, with the result that a larger part of the jet contributes to the integrated values of the polarization PA. In the case of jet bending, which is not so uncommon at low radio frequencies \citep{Kharb2010}, the resulting integrated value of the EVPA can easily be nonrepresentative of the jet position angle as inferred at higher frequencies where the bending is less severe.

\subsection{Jet direction versus radio EVPA at 86 GHz} \label{subsec:comp_86ghz}

\begin{figure}
   \centering
   \includegraphics[width=0.35\textwidth]{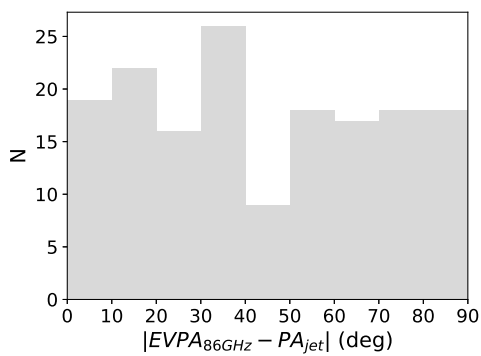}
      \caption{Distribution of the difference between the jet direction and EVPA at 86 GHz.}
         \label{fig:jd_vs_86GHz}
\end{figure}

We cross-matched our sample with the catalog of AGN polarization at 86 GHz from \cite{Agudo2014}, which resulted in 162 common sources. The distribution of the difference between PA$_{\rm jet}$ and the EVPA at 86 GHz is shown in Fig.~\ref{fig:jd_vs_86GHz}.
It has no significant peak and cannot be distinguished from a uniform distribution ($p$-value=0.17). This is in agreement with findings of \cite{Agudo2014}, who constructed the same distribution using PA$_{\rm jet}$ from the literature. As discussed in \cite{Agudo2014}, the absence of a peak at $0\dg$ or $90\dg$ contradicts theoretical predictions that EVPA must be either parallel or perpendicular to the jet axis in most of the cases for axisymmetric jets \citep[e.g.,][]{Lyutikov2005}. Several effects can smear out the expected peaks of the distribution in Fig.~\ref{fig:jd_vs_86GHz}. First, AGN are more variable in the total flux and polarization at higher frequencies \citep{Agudo2017}. This can largely blur the significant peak seen at 8 GHz in Fig.~\ref{fig:jd_vs_8ghz} by the much stronger variability at 86 GHz. Second, the emission at 86 GHz mostly comes from the radio core, which might be produced by an oblique shock in the jet. In this case, the EVPA does not necessarily have to be aligned or be perpendicular to the jet axis, but can vary in a wide range of angles depending on the system parameters \citep{Cawthorne2006}. Finally, the spectral indices at 86~GHz are highly variable and in some cases are consistent with transitions from optically thin to optically thick regimes \citep{Agudo2017}. These transitions can in turn be accompanied by $90\dg$ swings in EVPA \citep{Pacholczyk1970,Myserlis2016} that can contribute to spread of the $|EVPA_{\rm 86\,GHz}-PA_{\rm jet}|$ distribution.

\subsection{Jet direction versus optical EVPA} \label{subsec:comp_opt}

\begin{figure}
   \centering
   \includegraphics[width=0.35\textwidth]{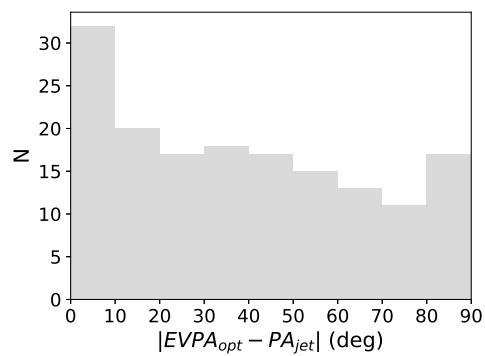}
      \caption{Distribution of the difference between the jet direction and optical EVPA.}
         \label{fig:jd_vs_opt}
\end{figure}

For the optical polarization data we repeated the same cross-matching procedure as for the radio. We
found 160 sources with measured optical EVPA and determined PA$_{\rm jet}$ from our sample. The 
distribution of the difference between the two angles is shown in Fig.~\ref{fig:jd_vs_opt}. 
It shows a significant peak at $0\dg$, which is consistent with an emission of an optically thin jet
with a toroidal magnetic field component dominating or at least not weaker than the poloidal 
component \citep{Lyutikov2005}. However, only 28\% of sources have $|EVPA_{\rm opt}-PA_{\rm jet}| < 
15\dg$, while the rest of the sample shows a rather uniform distribution with a possible minor peak 
at $90\dg$. Various effects jumble and flatten the distribution. For instance, optical and radio 
emission may probe separate regions with a different local magnetic field and jet orientation. In 
some sources optical emission may have a comparable contribution from the accretion disk and the 
jet. The polarization signatures of these two emitting components are drastically different and may 
introduce modifications in the orientation of the net emission EVPA with respect to the jet axis. 
Additionally, many of the sources included in the analysis are blazars from \cite{Angelakis2016} and 
\cite{Itoh2016}, which are known to be highly variable in flux, polarization degree, and 
polarization angle. For some of these sources only a few measurements are available. This means that 
the EVPA in these measurements could be far from its most probable value for a given blazar. 
Moreover, \cite{Angelakis2016} demonstrated that low synchrotron peaked ($\rm \nu_{peak} < 10^{13} 
Hz$) sources tend to have a uniform distribution of EVPA in monitoring data. The optical EVPA in 
these sources therefore does not carry any information on orientation and should not be used in 
alignments studies. On the other hand, high synchrotron peaked ($\rm \nu_{peak} > 10^{15} Hz$) 
sources tend to have stable EVPA concentrated near a preferred value \citep{Hovatta2016}.

We conclude that EVPA of emission at different frequencies is a rather weak indicator of the structural axis orientation of the AGN. Moreover, at some frequencies (e.g., 86 GHz) the polarization plane direction and AGN axis appear to be totally uncorrelated.

\section{Global alignments of jet directions} \label{sec:pa_align}

A variety of methods have been proposed to characterize the alignments of polarization planes \citep[for a review, see][]{Pelgrims2016b}. \cite{Jain2004} introduced a coordinate invariant statistics that eliminates the dependence of statistical tests and interpretation on a particular coordinate system. Because their method was applied to various data sets, including polarization data in different bands \cite{Jain2004,Tiwari2013,Tiwari2019} and data on the structural axes of the radio sources \cite{Contigiani2017}, we used the same method in order to be able to make direct comparisons with previous works.

Briefly, the method is as follows. 1) For every two sources $s_1$ and $s_2$ in our sample, we calculated three quantities: the angular separation between $s_1$ and $s_2$, the angle $\zeta_1$ between directions to $s_2$ and to the north pole from the position of $s_1$, and the angle $\zeta_2$ between directions to $s_1$ and to the north pole from the position of $s_2$; 2) For a given $i$-th source we defined the $n$ closest neighbors (including the $i$-th source itself) using the angular separations and calculated the following quantity:
\begin{equation}
d_{i,n}\biggr\rvert_{max} = \frac{1}{n} \left[ \left(\sum_{k=1}^n \cos{2 \alpha_k^\prime} \right)^2 + \left(\sum_{k=1}^n \sin{2 \alpha_k^\prime} \right)^2 \right]^{1/2},
\end{equation}
where $\alpha_k^\prime = \alpha_k + \zeta_i - \zeta_k$ is the jet position angle of the $k$-th source after it is transported to the location of the $i$-th source, while $\alpha_k$ is the original PA$_{\rm jet}$ of the $k$-th source at its initial location. The position angle changes due to the procedure of parallel transport, which is essential for the method and guarantees the coordinate invariance of the result. $d_{i,n}\rvert_{max}$ represents the maximized EVPA dispersion among $n$ neighbors at the $i$-th source position \citep{Contigiani2017}. For the sample of $N$ sources we used the following statistics, which defines the average maximized dispersion of EVPA among the $n$ closest neighbors in the sample:
\begin{equation}
    S_{n}=\frac{1}{N}\sum_{i=1}^nd_{i,n} \biggr\rvert_{max}.
    \label{eq:Sn}
\end{equation}
For $N\gg n\gg1,$ the distribution of $S_{n}$ is normal with the variance
\begin{equation}
    \sigma^{2}=\frac{0.33}{N},
    \label{eq:sigma2}
\end{equation}
as reported by \cite{Jain2004}. Following \cite{Contigiani2017}, we use the following significance level (S.L.) of the $S_{n}$ statistics:
\begin{equation}
    S.L.= 1-\Phi \left(\frac{S_{n}-\langle S_{n} \rangle _{MC}}{\sigma_{n}}\right),
\end{equation}
where $\Phi$ is the normal cumulative  distribution function. $\langle S_{n} \rangle _{MC}$ is the value found for $S_{n}$ through Monte Carlo (MC) simulations, assuming no alignments. Values of $\log(S.L.)=-1.3$ and $-2.5$ roughly correspond to $2\sigma$ and  $3\sigma$ confidence levels, respectively. A value of $\log(S.L.)<-2.5$ would be more than $3\sigma$ confidence. For a more detailed and mathematically accurate explanation of the method we refer to \cite{Jain2004} and \cite{Contigiani2017}.

In order to address the significance of PA$_{\rm jet}$ alignments, we conducted the MC simulations in two different ways. In the first approach, we shuffled the PA$_{\rm jet}$ values within the sample, that is, every source was assigned a value of another random source of our sample. After shuffling we computed $S_{n}$ for a given $n$. Then we repeated the procedure $10^4$ times for each $n$, and used the average $\langle S_{n} \rangle$ derived from the simulation. This shuffling process was used in similar studies to generate random data \citep[e.g.,][]{Hutsemekers1998,Hutsemekers2001,Jain2004}. In the second approach, we used the procedure described above, but instead of shuffling PA$_{\rm jet}$ values, we assigned a random value to each member of the sample that was uniformly distributed between $0\dg$ and $180\dg$.

\begin{figure}
   \centering
   \includegraphics[width=0.45\textwidth]{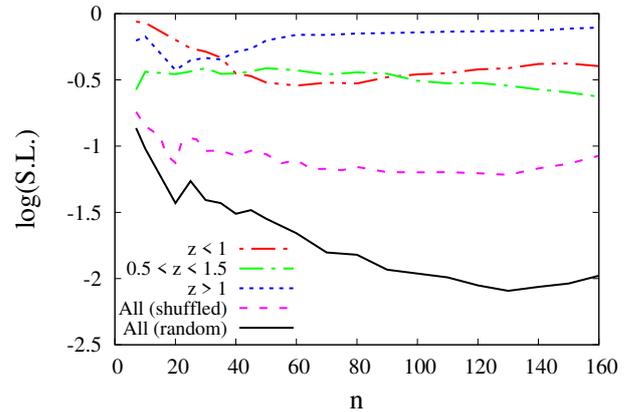}
      \caption{Significance level (S.L.) of the $S_n$ statistics as a function of $n$ the number of neighbors for different samples (see text for details).}
         \label{fig:S_n}
\end{figure}

To account for the possible dependence of the large-scale alignments on the redshift  \citep{Hutsemekers2005,Pelgrims2016}, we performed our calculations for the entire sample of PA$_{\rm jet}$ and for three $z$ ranges: $z<1$, $0.5<z<1.5$, and $z>1$. The redshift estimates for sources in our sample were obtained from the Optical Characteristics of Astrometric Radio Sources (OCARS) catalog\footnote{\url{http://www.gaoran.ru/english/as/ac_vlbi/ocars.txt}}. For all three redshift-cut subsamples we used the second method of MC simulations with uniformly distributed mock PA$_{\rm jet}$ to estimate the significance.

\begin{figure}
   \centering
   \includegraphics[width=0.40\textwidth]{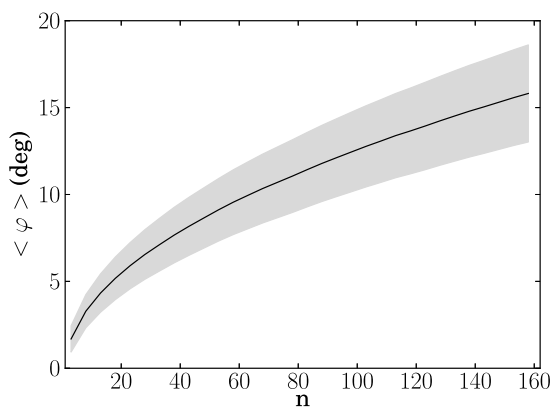}
      \caption{Median of the radii including $n$ closest neighbors as a function of $n$. The gray area represents the $1\sigma$ deviation from the median.}
         \label{fig:phi_n}
\end{figure}

The analysis results are shown in in Fig. \ref{fig:S_n}. In all cases the significance of the PA$_{\rm jet}$ alignments is found to be lower than the $3\sigma$ level. The dependence of the median radius $\left<\phi\right>$ of a circle on the sky including $n$ closest neighbors from our sample as a function of $n$ is shown in Fig.~\ref{fig:phi_n}. It follows from this figure that the characteristic angular scales that we probe in the analysis are between $\left<\phi\right>=2\dg$ and $15\dg$.

\section{Discussion and conclusions} \label{sec:concl}

The results of our comparison of EVPA at different frequencies with PA$_{\rm jet}$ in Sect.~\ref{sec:pa_vs_evpa} demonstrate that in general, the direction of the polarization plane at some frequencies can be used as a very weak proxy of the structural axis orientation of an AGN. However, at some frequencies (e.g., 86 GHz), the correlation between EVPA and PA$_{\rm jet}$ appears to be totally smeared out. This can explain, for example, why \cite{Tiwari2019} did not find any alignments of the EVPA at 86 GHz, while in other studies such alignments have been found at both radio and optical wavelengths \citep[e.g.,][]{Hutsemekers2005,Pelgrims2016,Tiwari2013}. Nevertheless, our results demonstrate that the correlations between EVPA and PA$_{\rm jet}$ are rather weak and are presumably strongly contaminated by various effects that alter both quantities. Because in only a small fraction (30-40\%) of sources the EVPA is either parallel or perpendicular to the jet axis within the uncertainties, it is rather surprising that multiple studies did find significant large-scale alignments of the EVPA. Such studies can be improved by considering the effects that may be imposing the EVPA. For instance, the Faraday rotation at low radio frequencies can be partially accounted for with observations in multiple bands. On the other hand, studies of optical EVPA alignments should avoid using low synchrotron peaked sources that have highly variable EVPA with a uniform distribution \citep{Angelakis2016}. 
Moreover, the significance of the results of such studies in optical bands could be drastically improved if only sources with a strongly dominating accretion disk or jet emission were considered. For instance, recently discovered correlation between optical polarization parameters and VLBI-Gaia offsets of AGN positions (Kovalev et al., accepted) offers a tool for selection of such sources.

\begin{figure}
   \centering
   \includegraphics[width=0.4\textwidth]{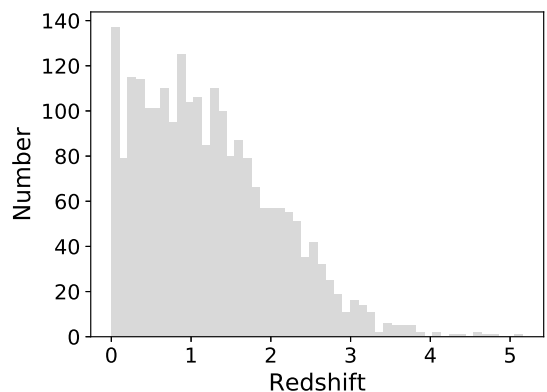}
      \caption{Redshift distribution for the sample sources.}
         \label{fig:z_dist}
\end{figure}

The results of our analysis in section~\ref{sec:pa_align} imply that PA$_{\rm jet}$ does not show significant global alignments at angular scales between $2\dg$ and $15\dg$  for the entire sample or in different redshift bins. The redshift distribution in our sample is shown in Fig.~\ref{fig:z_dist}.
The median value of $z=1.1$, which scales the lowest end of the probed angular sizes $\left<\phi\right>=2\dg$ to an average linear size of 60.5 Mpc.

\cite{Contigiani2017} performed very similar analysis testing the alignments of long axes of radio sources. They found significant alignments at linear scales between 19 and 39 Mpc, which 
corresponds to $\left<\phi\right>=1.5\dg$ for the average $z=0.47$ in their sample. Therefore, the results of our study complement the work of \cite{Contigiani2017} and confirm their results, showing that at linear scales $>40$ Mpc, radio sources do not show alignment of their structural axes. However, our study covers a different jet scale than \cite{Contigiani2017}. Because radio jets at parsec and kilo-parsec scales are often misaligned \citep[e.g.,][]{Pearson1988,Conway1993,Appl1996,Kharb2010}, it is not clear whether these two studies can be directly compared.

\begin{acknowledgements}
      We thank L. Petrov who maintains the Astrogeo archive and all contributors to this rich 
dataset. We thank Z. Malkin who maintains the OCARS database. We thank S. Savchenko and A. Roy for 
helpful comments. D.B. and N.M. acknowledge support from the European Research Council (ERC) under 
the European Union Horizon 2020 research and innovation program under the grant agreement No 771282.
\end{acknowledgements}

%
%
\bibliographystyle{aa}
\bibliography{bibliography}

\end{document}